\documentclass[aps,prd,twocolumn,superscriptaddress,showpacs,preprintnumbers,amsmath,amssymb]{revtex4-1}
\usepackage{dcolumn}  

\usepackage{amsmath}
\usepackage{amssymb}
\usepackage{bm}
\usepackage{expl3}
\usepackage{graphicx}
\usepackage[italic]{hepnames}
\usepackage{multirow}
\usepackage{physics}
\usepackage[separate-uncertainty=true,retain-explicit-plus,group-separator={}]{siunitx}
\usepackage{wasysym}
\usepackage{xcolor}
\usepackage{xfp}
\usepackage{xfrac}
\usepackage{xpatch}
\usepackage{dcolumn}  

\makeatletter
\xpatchcmd\@HepConStyle
 {\edef\@upcode{\updefault}}
 {\ifdefined\shapedefault\edef\@upcode{\shapedefault}\else\edef\@upcode{\updefault}\fi}
 {}{}
\makeatother

\newcommand{\Sup}[1]{\ensuremath{^{\text{#1}}}\xspace}

\newcommand{\roundEverything}[1]{\sisetup{round-mode=places,round-precision=#1}}

\newcommand{\roundednum}[2]{\num[round-mode=places,round-precision=#1]{#2}}
\newcommand{\roundedSI}[4]{\ifx&#4&\SI[round-mode=places,round-precision=#1]{#2}{#3}\else$(\roundednum{#1}{#2} \pm \roundednum{#1}{#3})$\;\!\si{#4}\fi}

\renewcommand{\Ppizero}{\ensuremath{\Ppi^0}\xspace}
\newcommand{\PKstarzero}{\ensuremath{\PK^{*0}}\xspace}
\newcommand{\PKstarplus}{\HepParticle{K}{}{*+}\xspace}
\newcommand{\PKstarminus}{\HepParticle{K}{}{*-}\xspace}
\newcommand{\APKstar}{\HepAntiParticle{K}{}{*}\xspace}
\newcommand{\APKstarzero}{\ensuremath{\overline{K}^{*0}}\xspace}
\renewcommand{\PUpsilonOneS}{\HepParticleResonance{\PUpsilon}{\text{1S}}{}{}\xspace}

\newcommand{\PUpsilonFiveS}{\HepParticleResonance{\PUpsilon}{\text{5S}}{}{}\xspace}
\renewcommand{\PKshort}{\ensuremath{\PK^0_{\text{S}}}\xspace}
\renewcommand{\PKlong}{\ensuremath{\PK^0_{\text{L}}}\xspace}

\renewcommand{\PKzero}{\ensuremath{\PK^0}\xspace}
\renewcommand{\PDzero}{\ensuremath{\PD^0}\xspace}

\newcommand{\KKpp}{\ensuremath{{\PKminus\!\PKshort\;\!\Ppiplus\!\Ppiplus}}\xspace}
\newcommand{\KKppp}{\ensuremath{{\KKpp\!\Ppizero}}\xspace}
\newcommand{\Kppppp}{\ensuremath{{\PKminus\!\Ppiplus\!\Ppiplus\!\Ppiplus\!\Ppiminus\!\Ppizero}}\xspace}
\newcommand{\Kep}{\ensuremath{{\APKstar\!\!\;\Peta\!\;\Ppiplus}}\xspace}
\newcommand{\Kop}{\ensuremath{{\APKstar\!\!\;\Pomega\!\;\Ppiplus}}\xspace}
\newcommand{\DtoKKppp}{\ensuremath{{\PDplus \!\! \to \KKppp}}\xspace}
\newcommand{\DtoKKpp}{\ensuremath{{\PDplus \!\! \to \KKpp}}\xspace}
\newcommand{\DtoKppppp}{\ensuremath{{\PDplus \!\! \to \Kppppp}}\xspace}
\newcommand{\DtoKep}{\ensuremath{{\PDplus \!\! \to \Kep}}\xspace}
\newcommand{\DtoKop}{\ensuremath{{\PDplus \!\! \to \Kop}}\xspace}

\begin{document}


\title{Search for \DtoKKppp at Belle}

%

\collaboration{The Belle Collaboration}

\noaffiliation
\date{\today}

\def\SystUncEffUp{12.6038}
\def\SystUncEffDown{6.4428}
\def\SystUncPiZ{2.3}
\def\SystUncUp{\fpeval{sqrt(\SystUncEffUp^2 + \SystUncPiZ^2)}}
\def\SystUncDown{\fpeval{sqrt(\SystUncEffDown^2 + \SystUncPiZ^2)}}
\def\BFFourPartUnc{\fpeval{0.17 / 2.27 * 100}}
\def\BFKKpiOrder{e-5}
\def\BFKKpiMode{6.353}
\def\BFKKpiStatUnc{3.778}
\def\BFKKpiSignificance{1.68908}
\def\BFKKpiUpperLimit{1.36396e-4}
\def\EffKKpi{4.50}              
\def\EffKKpiUnc{0.0168}         
\def\VetoEffKKpi{76.339}        
\def\VetoEffKKpiUnc{0.4034}     
\def\NKKpi{8474}
\def\fracKKpi{0.0181}
\def\fracKKpiUnc{0.0082843}
\def\YKKpi{\fpeval{\NKKpi * \fracKKpi}}
\def\YKKpiUnc{\fpeval{\NKKpi * \fracKKpiUnc}}
\def\NKKpiWithoutVeto{14682}
\def\fracKKpiWithoutVeto{0.026388}
\def\fracKKpiWithoutVetoUnc{0.0067257}
\def\YKKpiWithoutVeto{\fpeval{\NKKpiWithoutVeto * \fracKKpiWithoutVeto}}
\def\YKKpiWithoutVetoUnc{\fpeval{\NKKpiWithoutVeto * \fracKKpiWithoutVetoUnc}}
\def\EffNorm{12.9822}           
\def\EffNormUnc{0.03019}        
\def\NNorm{5929690}
\def\fracNorm{0.0238788}
\def\fracNormUnc{0.000240931}
\def\YNorm{\fpeval{\NNorm * \fracNorm}}
\def\YNormUnc{\fpeval{\NNorm * \fracNormUnc}}
\def\YFive{136.615}
\def\YFiveUnc{81.8702}
\def\BFeoOrder{e-5}
\def\BFeoMode{6.01831}           
\def\BFeoStatUncUp{6.23941}      
\def\BFeoStatUncDown{5.74372}    
\def\BFeoUpperLimit{1.7561e-4}  
\def\BFeoSignificance{1.09243}   
\def\Effeo{5.27243}             
\def\EffeoUnc{0.6}              
\def\VetoSupeo{90.399}          
\def\VetoSupeoUnc{1.0745}       
\def\BFSumOrder{e-4}
\def\BFSumMode{1.28862}          
\def\BFSumStatUncUp{0.372496}    
\def\BFSumStatUncDown{0.342519}  
\def\BFSumUpperLimit{2.13178e-4} 
\def\BFSumSignificance{3.47012}  
\def\BFsigBFeoStatCorr{-0.797376}   
\def\BFsigBFSumStatCorr{-0.244075}  
\def\BFeoBFSumStatCorr{0.779576}      
\def\BFKKpiSystUncUp{\fpeval{\BFKKpiMode * \SystUncUp / 100}}
\def\BFKKpiSystUncDown{\fpeval{\BFKKpiMode * \SystUncDown / 100}}
\def\BFKKpiFourPartUnc{\fpeval{\BFKKpiMode * \BFFourPartUnc / 100}}
\def\BFeoSystUncUp{\fpeval{\BFeoMode * \SystUncUp / 100}}
\def\BFeoSystUncDown{\fpeval{\BFeoMode * \SystUncDown / 100}}
\def\BFeoFourPartUnc{\fpeval{\BFeoMode * \BFFourPartUnc / 100}}
\def\BFSumSystUncUp{\fpeval{\BFSumMode * \SystUncUp / 100}}
\def\BFSumSystUncDown{\fpeval{\BFSumMode * \SystUncDown / 100}}
\def\BFSumFourPartUnc{\fpeval{\BFSumMode * \BFFourPartUnc / 100}}
\def\BFKKpiFullUncUp{\fpeval{sqrt(\BFKKpiStatUnc^2 + \BFKKpiSystUncUp^2 + \BFKKpiFourPartUnc^2)}}
\def\BFKKpiFullUncDown{\fpeval{sqrt(\BFKKpiStatUnc^2 + \BFKKpiSystUncDown^2 + \BFKKpiFourPartUnc^2)}}
\def\BFeoFullUncUp{\fpeval{sqrt(\BFeoStatUncUp^2 + \BFeoSystUncUp^2 + \BFeoFourPartUnc^2)}}
\def\BFeoFullUncDown{\fpeval{sqrt(\BFeoStatUncDown^2 + \BFeoSystUncDown^2 + \BFeoFourPartUnc^2)}}
\def\BFSumFullUncUp{\fpeval{sqrt(\BFSumStatUncUp^2 + \BFSumSystUncUp^2 + \BFSumFourPartUnc^2)}}
\def\BFSumFullUncDown{\fpeval{sqrt(\BFSumStatUncDown^2 + \BFSumSystUncDown^2 + \BFSumFourPartUnc^2)}}

\begin{abstract}
  \roundEverything{1}

  We search for the singly Cabibbo-suppressed decay \DtoKKppp using
  the full data set of \SI{988}{fb^{-1}} recorded by the Belle
  experiment at the KEKB $\APelectron\Pelectron$ collider. We measure
  the branching fraction for this decay to be
  $(\num{\BFKKpiMode}^{\num{+\BFKKpiFullUncUp}}_{\num{-\BFKKpiFullUncDown}})\times\num{\BFKKpiOrder}$
  with an upper limit at 95\% credibility of
  \num{\BFKKpiUpperLimit}. We also measure the sum of the branching
  fractions for \DtoKep and \DtoKop to be ${(\num{\BFeoMode} \,
  ^{\num{+\BFeoFullUncUp}}_{\num{-\BFeoFullUncDown}}) \times
  \num{\BFeoOrder}}$ with an upper limit at 95\% credibility of
  \num{\BFeoUpperLimit}.

\end{abstract}

\maketitle

\section{Introduction}

The standard model of particle physics predicts little to no CP
violation in the decay of charmed mesons. Searching for
larger-than-expected asymmetries in these decays probes physics beyond
the standard model. The LHCb collaboration reported the first
observation of CP violation in charmed-meson decay in their
measurement of decays of \PDzero to $\PKplus\PKminus$ and
$\Ppiplus\Ppiminus$~\cite{Aaij:2019kcg}. This asymmetry is compatible
with many reported standard model expectations, but since perturbative
calculations are difficult at the scale of the charm mass, there is no
consensus that the standard model alone explains
it~\cite{Grossman:2006jg,Brod:2011re,lnguglia:2013mha,Cheng:2012wr,Bhattacharya:2012ah,Cheng:2019ggx,Charles:2011va,Altmannshofer:2012ur,Giudice:2012qq}.

The decays in which CP violation was observed involve changes of
isospin, $I$, by both $\frac12$ and $\frac32$. In the limit of SU(3)
flavor symmetry, the standard model allows for CP violation only in
$\Delta I{=}\frac12$
transitions~\cite{Grossman:2006jg,Grossman:2012eb}. The only pure
$\Delta I{=}\frac32$ charmed-meson decay is \HepProcess{\PDplus \to
  \Ppiplus \Ppizero}. In this decay, SU(3)-flavor-breaking effects
allow for CP violation in the standard model at less than
$\order*{\num{e-4}}$~\cite{Grossman:2012eb}. LHCb reported the most
precise measurement of the CP asymmetry in this decay:
\SI[parse-numbers=false]{(-1.3\pm1.1)}{\!\percent}~\cite{LHCb:2021rou}. This is consistent
with the standard model expectation but still allows for sizeable
beyond-standard-model effects.

One can isolate $\Delta I{=}\frac32$ transition amplitudes in
combinations of amplitudes for decays that also involve $\Delta
I{=}\frac12$ transitions. In decays of \PDplus to
$\PKstar\APKstar\Ppi$, we can isolate a particular $\Delta
I{=}\frac32$ amplitude---namely that to a final state with total
isospin 2 with the two kaons having total isospin 1---via the sum
\begin{equation}
  \mathcal{A}(\PKstarplus\PKstarminus\Ppiplus)
  + \mathcal{A}(\PKstarzero\APKstarzero\Ppiplus)
  + \sqrt2 \mathcal{A}(\PKstarplus\APKstarzero\Ppizero),
  \label{eqn:A32sum}
\end{equation}
where the amplitudes are for decays to the particular charge
configurations of $\PKstar\APKstar\Ppi$~\cite{Grossman:2012eb}. From
this amplitude sum, we can calculate the asymmetry in a $\Delta
I{=}\frac32$ transition. This requires we measure the relative
magnitudes and phases of the three amplitudes in the sum. We can do
this via the five-particle final state common to all three
$\PKstar\APKstar\Ppi$ charge configurations:
$\PKminus\PKzero\Ppiplus\Ppiplus\Ppizero$. This decay is
unobserved. We report the first search for \DtoKKppp and improve the
measurement of the branching fractions for \DtoKep and \DtoKop, which
are a potential source of background.

\section{Data selection}

We use the full data set recorded by the Belle experiment at the KEKB
asymmetric-energy $\APelectron\Pelectron$
collider~\cite{Abashian:2000cg, Brodzicka:2012jm, Kurokawa:2001nw,
  *Abe:2013kxa}. The center-of-momentum~(c.m) energy of collisions
varied from the mass of the \PUpsilonOneS resonance up to that of the
\PUpsilonFiveS resonance. The integrated luminosity of the data is
\SI{988}{fb^{-1}}~\cite{Brodzicka:2012jm}.

The Belle detector was a large-solid-angle magnetic spectrometer that
consisted of a silicon vertex detector~(SVD), a fifty-layer central
drift chamber~(CDC), an array of aerogel threshold Cherenkov counters,
a barrel-like arrangement of time-of-flight scintillation counters,
and an electromagnetic calorimeter~(ECL) comprised of CsI(Tl) crystals
located inside a superconducting solenoid coil that provided a
1.5-\si{T} magnetic field. An iron flux-return yoke located outside of
the coil was instrumented to detect \PKlong's and identify muons. A
more detailed description of the detector and its performance is found
in Refs.~\cite{Abashian:2000cg} and~\cite{Brodzicka:2012jm}.

We measure the branching fraction for the five-particle decay, \DtoKKppp,
relative to that of the four-particle decay, \DtoKKpp:
\begin{equation}
  \mathcal{B}(\KKppp)
  = \frac{Y_5 / \epsilon_5}{Y_4 / \epsilon_4} \times \mathcal{B}(\KKpp),
  \label{eqn:bf_calc}
\end{equation}
where $Y_n$ and $\epsilon_n$ are the yields of observed events
(determined from fits to data) and the detection efficiencies,
respectively, of the two decays where $n$ labels the number of
particles in the final state~\footnote{For brevity of notation, we
denote all branching fractions by the final state since the initial
state, \PDplus, is always the same. All results include the
charge-conjugated decays.}. The branching fraction for the four-particle
decay is~\num{2.27+-0.17e-3}~\cite{Link:2001hz, PDG2020}. Accounting
for the phase-space suppression of an additional \Ppizero, we expect
the branching fraction for the five-particle final state to be of
order~\num{e-4}.

To develop our event selection criteria without possible bias from
inspecting the data, we use simulated data of the four- and
five-particle decays in an amount equivalent to three times the Belle
data, with the final-state particles in both decays uniformly
distributed in the available phase space. To determine what types of
background events pass our selection criteria, we use simulated data
of \HepProcess{\APelectron\Pelectron \to \Pquark\APquark} in an amount
equivalent to five times the Belle data.

We use the above-described simulated signal data along with simulated
data of the three charge configurations of \HepProcess{\PDplus \to
  \PKstar\APKstar\Ppi} to determine the detection efficiency and
related systematic uncertainties. We generated \num{e7} events in each
$\PKstar\APKstar\Ppi$ charge configuration with the three-particle
state uniformly distributed in the available phase space. For all
simulated data, we modeled particle production and decay with
EvtGen~\cite{Lange:2001uf} and the Belle detector response with
GEANT3~\cite{Brun:1987ma}.

To select events in which charm mesons may be created, we consider
events in which the ratio of the second to the zeroth Fox-Wolfram
moment is above \num{0.1}~\cite{PhysRevLett.41.1581}. For the charged
decay products of the \PDplus, we only consider particles with two or
more hits in the SVD and distances of closest approach to the
interaction point of the \APelectron and \Pelectron beams (IP) below
\SI{15}{mm} in the longitudinal direction and \SI{10}{mm} in the
transverse plane (defined here and throughout with respect to the
positron direction) and only those inconsistent with being a lepton or
proton. We identify a particle as a pion if its likelihood to be a pion
(rather than a kaon) is greater than~\SI{60}{\!\percent}; otherwise we
identify it as a kaon.

We reconstruct neutral kaons from $\Ppiplus\Ppiminus$ pairs selected
by a neural network~\cite{FEINDT2006190} that considers the \PKshort
momentum in the lab frame, the distance between the \Ppiplus
and~\Ppiminus tracks in the longitudinal direction, the flight length
of the \PKshort in the transverse direction, the angle between the
\PKshort momentum and its displacement from the IP, the distances of
closest approach of the \Ppiplus and \Ppiminus, the angle between the
center-of-momentum system in the lab frame and the positive pion in
the \PKshort rest frame, and information about the pion hits in the
SVD and CDC. Each of the pions forming a \PKshort must have lab-frame
momentum above~\SI{60}{MeV}~\footnote{We use a unit system in whcih
  energy, mass, and momentum all have units of \si{eV}.}. Their
longitudinal separation must be less than~\SI{20}{cm}. And their
invariant mass must be within~\SI{20}{MeV} of the known \PKzero
mass~\cite{PDG2020}. We constrain the tracks of each selected pion
pair to have a common origin and invariant mass equal to the known
\PKzero mass. The \PKshort momenta are calculated from the constrained
tracks.

We reconstruct neutral pions from photon pairs, requiring each
\Ppizero candidate have a mass within \SI{20}{MeV} of the known
\Ppizero mass (the mass resolution is \SI{5}{MeV}) and momentum
greater than \SI{220}{MeV} in the $\APelectron\Pelectron$
center-of-momentum frame~\cite{PDG2020}. Photons are defined as
clusters in the ECL unassociated with any charged particle and with
more than \SI{85}{\!\percent} of their energy deposited in the
$3{\times}3$ grid of ECL crystals centered around the crystal with the
highest energy deposition, in comparison to the energy deposited in
the likewise centered $5{\times}5$ grid. We constrain the four-momenta
of each selected photon pair to have invariant mass equal to the known
\Ppizero mass. The \Ppizero momenta are calculated from the
constrained momenta.

For both the signal and normalization decays, we accept a set of
final-state particles as the decay products of a \PDplus candidate if
their invariant mass is within \SI{50}{MeV} of the known \PDplus
mass~\cite{PDG2020}. This window is significantly larger than the
\PDplus mass resolution in both decay channels. To veto random
combinations of final-state particles, which tend to have low total
momenta, we select only \PDplus candidates with normalized momenta
greater than \num{0.4}; the normalized momentum is
\begin{equation}
  x_{\PD} \equiv \abs{\vec{p}_{\PD}} \bigg/ \! \sqrt{\frac{s}{4} - m_{\PD}^2},
  \label{eqn:normalized_momentum}
\end{equation}
where $\vec{p}_{\PD}$ is the c.m.-frame \PDplus momentum, $\sqrt{s}$ the
c.m. energy, and $m_{\PD}$ the known \PDplus mass. The denominator is
the maximum possible momentum a \PDplus can have in any event, so
$x_{\PD}$ is bounded between zero and one. The $x_{\PD}$ distribution
for random combinations of final-state particles peaks sharply at
zero. The distributions for true \PDplus produced in $\Pcharm\APcharm$
and $\Pbottom\APbottom$ events are broad and peak well above zero.

For each candidate passing the above criteria, we constrain the
charged final-state particles and \PKshort to have a common origin. We
reject candidates for which the constraining fit fails to converge. In
4\% of events, we find multiple \PDplus candidates. In such events, we
accept only the candidate with the highest vertex fit
probability. This selects the true decay with 76\% probability.

To further suppress background events in the five-particle decay, we
train a boosted decision tree~(BDT) on simulated events using
thirty-one kinematic variables of the \PDplus candidates and their
decay products~\cite{hoecker2007tmva,Levit:2019jqe}. The variables
with the highest distinguishing power are the length of the projection
of the \PDplus displacement in the transverse plane, the energy of the
\Ppizero in the rest frame of the \PDplus, and the transverse
momentum, normalized momentum, and vertex-fit $p$ value of the
\PDplus. We accept events with an output from the BDT above a value
that maximizes the signal yield divided by the square root of the
background yield (in simulated data). This rejects~98\% of background
events and accepts 26\% of signal events.  To check for over training
of the BDT, we repeat the efficiency calculation with several
independent sets of simulated data. The variation of the efficiency is
consistent within its statistical uncertainty.

\section{Branching fraction fit}

Three types of events pass our selection criteria: those containing
our signal decays, those containing random combinations of particles
that appear to form a \PDplus (background), and for the five-particle
decay, those containing \DtoKppppp in which no true \PKshort is
present. Since these events have already passed our \PKshort selection
criteria, they differ from our signal only by their potential origin
from \DtoKep and \DtoKop~\cite{Barlag:1992ww}. We veto such decays by
rejecting any \PDplus candidate for which any combination of
$\Ppiplus\Ppiminus\Ppizero$ has an invariant mass within a window
around the known \Peta or \Pomega masses~\cite{PDG2020} chosen to
contain 90\% of such events. This retains
\roundedSI{1}{\VetoEffKKpi}{\VetoEffKKpiUnc}{\!\percent} of signal
events and removes
\roundedSI{1}{\VetoSupeo}{\VetoSupeoUnc}{\!\percent} of \Kep and \Kop
events. The uncertainties on all efficiencies arise from the sizes of
the simulated data sets used to estimate them.

After all the above criteria, the signal selection efficiencies are
\roundedSI{2}{\EffKKpi}{\EffKKpiUnc}{\!\permil} for the five-particle
decay and \roundedSI{2}{\EffNorm}{\EffNormUnc}{\!\percent} for the
four-particle decay. To determine both signal fractions, we perform
unbinned maximum-posterior fits and sample the full posterior using
the Bayesian Analysis
Toolkit~\cite{caldwell2009bat,beaujean_frederik_2018_1322675}. In the
fits, we parameterize the signal and background components of our
likelihood as functions of the invariant mass of the \PDplus
candidate, whose shapes are determined from studying simulated events.

For each channel, we model the distribution of true \PDplus events as
a weighted sum of a double-sided Crystal-Ball (DCB) distribution and
two normal distributions~\cite{Skwarnicki:1986xj}. The signal
distribution has twelve free parameters: six for the DCB distribution,
two for each of the normal distributions, and two for the weights of
the normal distributions with respect to the DCB distribution. For
each channel, we model the distribution of background events as a
second-order polynomial with two free parameters. In total, each fit
has fifteen free parameters: fourteen shape parameters and one for the
fraction of signal events in the data.

In the fit to the experimental data, we set the prior probability
distribution of each shape parameter to the posterior probability
distributions obtained from a fit to simulated data. We use flat
priors for the signal-fraction parameters.

\begin{figure}[t]
  \centering
  \includegraphics[width=0.48\textwidth]{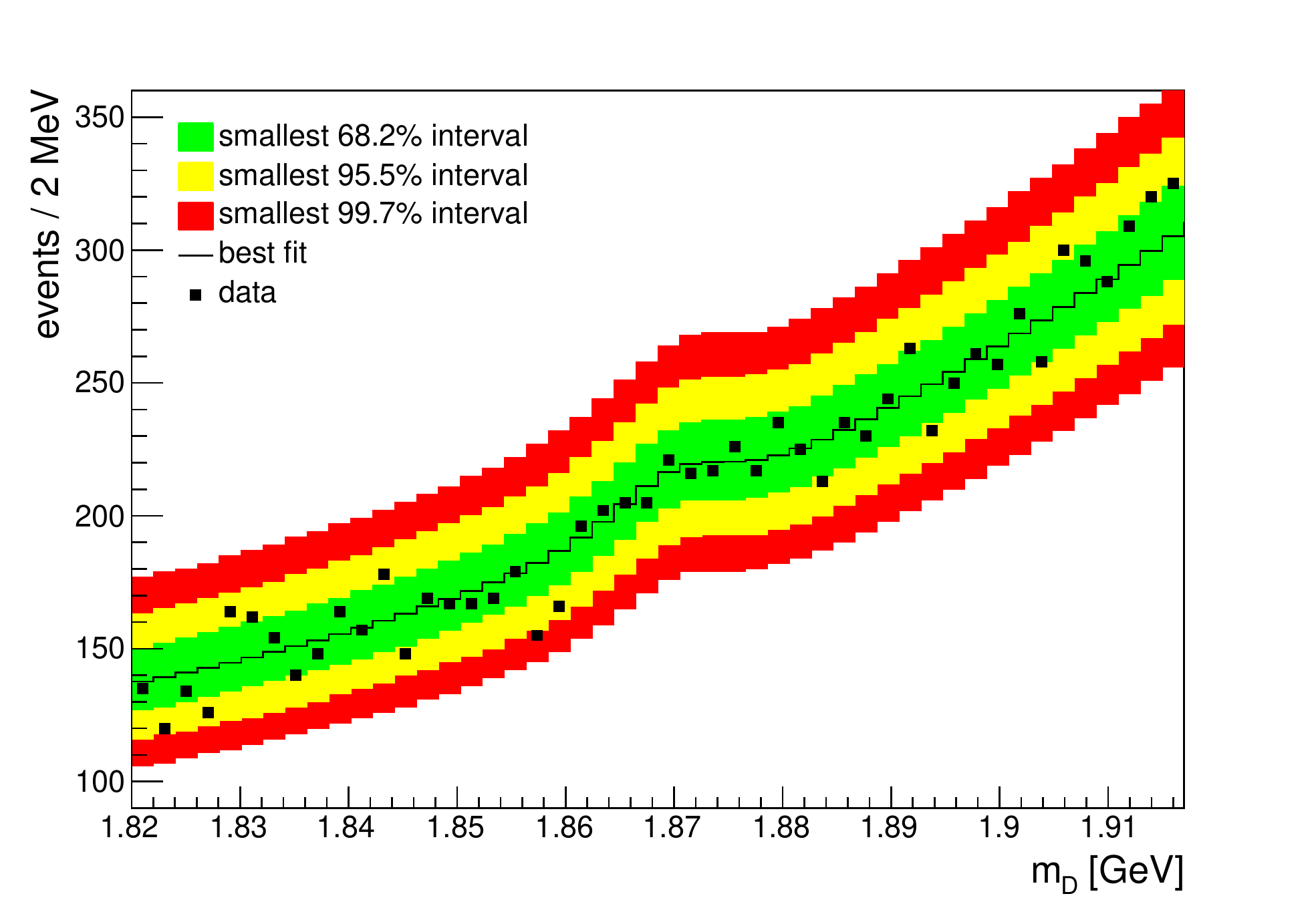}
  \hfill
  \includegraphics[width=0.48\textwidth]{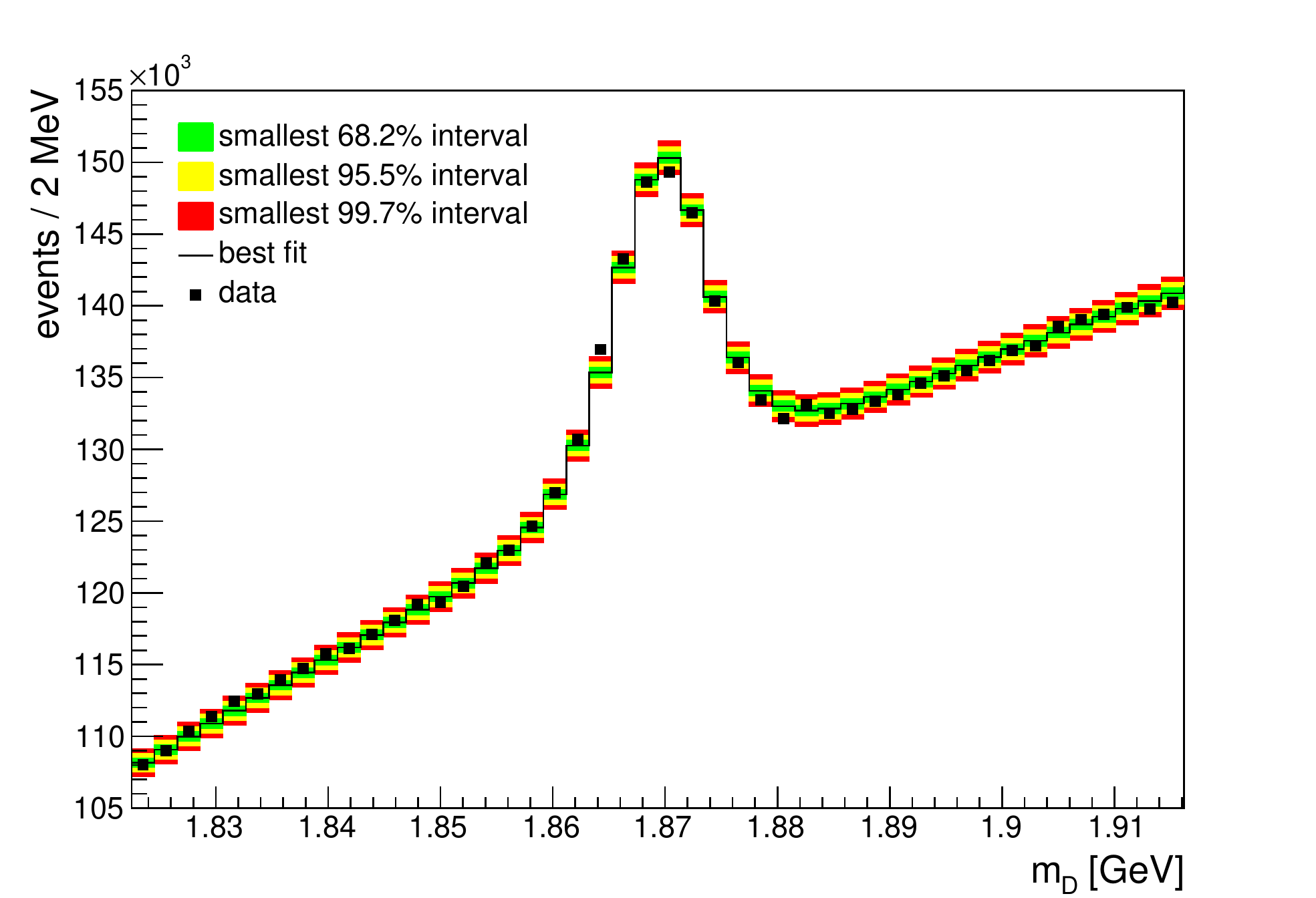}
  \caption{Invariant mass distributions of \PDplus candidates in
    \DtoKKppp (top) and \DtoKKpp (bottom) of the data (dots). The
    results of the unbinned fits are projected into the same bins as
    the data at the best fit point (line) and as bands of posterior
    probability for the observed number of events. The vertical axes
    start from nonzero values to highlight the shapes of the data and
    models.
    \label{fig:data_and_fits}
  }
\end{figure}

Figure~\ref{fig:data_and_fits} shows the invariant-mass distributions
of the data for the five- and four-particle decays with the results of
the fits shown as bands of posterior probability corresponding to the
typically given one-, two-, and three-standard-deviation intervals. We
see that the fits describe the data well. They determine the signal
yields to be \roundedSI{0}{\YKKpi}{\YKKpiUnc}{events} for the
five-particle decay and \roundedSI{0}{\YNorm}{\YNormUnc}{events} for
the four-particle decay.  The uncertainties are statistical only.

\begin{figure}[t]
  \centering
  \includegraphics[width=0.5\textwidth]{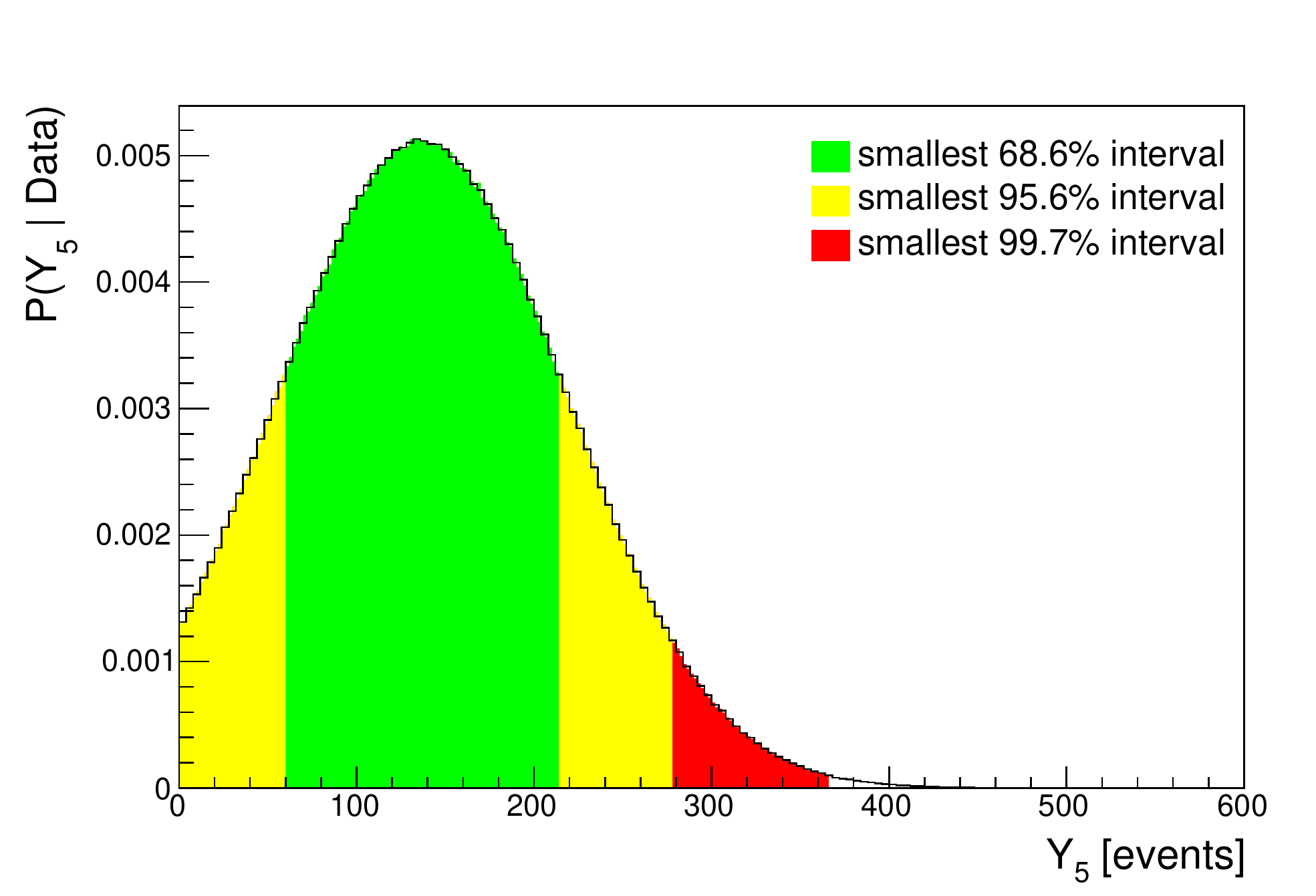}
  \caption{Posterior probability distribution for the true
    five-particle signal yield. \label{fig:dto5p_f}
  }
\end{figure}

After the \Peta and \Pomega vetoes, the five-particle signal-like
yield still contains some $\APKstarzero\Peta\Ppiplus$ and
$\APKstarzero\Pomega\Ppiplus$ events. Although we can precisely
estimate our suppression of this component, we cannot estimate its
size after suppression since the branching fractions for decay to \Kep
or \Kop are poorly measured~\cite{Barlag:1992ww}. Using the signal
yield from a fit without the \Peta and \Pomega vetoes, $y$, we can
determine $Y_5$, which is needed for the branching fraction
determination:
\begin{equation}
  Y_5 = \frac{\epsilon_5\Sup{v}}{\epsilon_5\Sup{v} - \epsilon_{\Peta\Pomega}\Sup{v}} \qty(Y - \epsilon_{\Peta\Pomega}\Sup{v} y),
  \label{eqn:Y5}
\end{equation}
where $Y$ is the signal yield with the vetoes, given above, and
$\epsilon_5\Sup{v}$ and $\epsilon_{\Peta\Pomega}\Sup{v}$ are the
efficiencies of the vetoes (given earlier) for the signal and the
decays to \Kep or \Kop determined from simulated data. The fit without
the vetoes yields
\roundedSI{0}{\YKKpiWithoutVeto}{\YKKpiWithoutVetoUnc}{events}.  The
signal yield $Y_5$ (in the fit with vetoes) is
\roundedSI{0}{\YFive}{\YFiveUnc}{events}. Figure~\ref{fig:dto5p_f}
shows the posterior probability distribution for~$Y_5$. The posterior
is normal, but cut off by its physical lower limit, zero.

Since we calculate the branching fraction from the ratio
$\flatfrac{Y_5}{Y_4}$, many systematic uncertainties related to
particle detection and identification cancel to negligible values. The
systematic uncertainty on the relative branching fraction comes from
two sources: the detection of the \Ppizero in the five-particle decay
and the estimation of the efficiency.

The systematic uncertainty arising from $\Ppizero$ detection has been
studied in \HepProcess{\Ptauon \to \Ppiminus\Ppizero\Pnut}.  For
neutral pions in our momentum range (near or below \SI{500}{MeV}), the
relative systematic uncertainty is \roundedSI{1}{\SystUncPiZ}{\!\percent}{}.

The efficiency for detecting the five particle state depends greatly
on where events are distributed in the phase space available to the
decay. This distribution depends on what intermediate resonances (and
what spin configurations thereof) there are between the \PDplus and
the final state. Since the efficiency is most sensitive to the
\Ppizero momentum, we map the efficiency (from simulated data) in the
two-dimensional plane of the squared invariant mass of the
$\Ppiplus\Ppiplus\Ppizero$ system versus the squared invariant mass of
the $\PKminus\PKshort\Ppizero$ system. The efficiency is calculated in
bins of this plane from simulated data in which the five final-state
particles are uniformly distributed in the available phase space. The
efficiency used to calculate the branching fraction is a sum of the
values in these bins weighted by the distribution of events evenly
distributed in phase space.

To calculate the systematic uncertainty arising from nature having a
different distribution in phase space than what we use to calculate
the efficiency, we calculate the efficiency for three further models
of the decay: via $\PKstarplus\PKstarminus\Ppiplus$, via
$\PKstarzero\APKstarzero\Ppiplus$, and via
$\PKstarplus\APKstarzero\Ppizero$, with the three particles of each
model distributed uniformly in the available (three-particle) phase
space. We take the spread of the efficiencies as a systematic
uncertainty. The resulting relative systematic uncertainty is
asymmetric: \roundedSI{1}{+\SystUncEffUp}{\!\percent}{} and
\roundedSI{1}{-\SystUncEffDown}{\!\percent}{}.

We sum both systematic uncertainties in quadrature, yielding a total
systematic uncertainty of \roundedSI{1}{+\SystUncUp}{\!\percent}{}
and~\roundedSI{1}{-\SystUncDown}{\!\percent}{}.

The above values yield a branching fraction for the five-particle decay
\begin{equation}
  \roundEverything{1}
  \mathcal{B}(\KKppp)
  = \qty(\num{\BFKKpiMode} \, \pm \num{\BFKKpiStatUnc} \, ^{\num{+\BFKKpiSystUncUp}}_{\num{-\BFKKpiSystUncDown}} \pm \num{\BFKKpiFourPartUnc}) \times \num{\BFKKpiOrder},
  \label{eqn:B5}
\end{equation}
where the first uncertainty is statistical, the second is systematic,
and the third is from the uncertainty on the four-particle branching
fraction~\cite{PDG2020}. Compared to the hypothesis of there being no
five-particle decay, this result has a significance of
\roundednum{1}{\BFKKpiSignificance} standard deviations. 

Equation~(\ref{eqn:Y5}), along with $Y = Y_5 + Y_{\Peta\Pomega}$,
gives us the combined yield for decay to \Kep or \Kop in our data. We
can calculate the sum of branching fractions for both decays:
\begin{equation}
  \roundEverything{1}
  \mathcal{B}(\Kep)
  +
  \mathcal{B}(\Kop)
  = \qty(\num{\BFeoMode} \, ^{\num{+\BFeoStatUncUp}}_{\num{-\BFeoStatUncDown}} \, ^{\num{+\BFeoSystUncUp}}_{\num{-\BFeoSystUncDown}} \pm \num{\BFeoFourPartUnc}) \times \num{\BFeoOrder}
  \label{eqn:B6}
\end{equation}
where we have used the (combined) efficiency for these channels,
\roundedSI{1}{\Effeo}{\EffeoUnc}\times{\num{e-5}}. The uncertainties
are as given in equation~(\ref{eqn:B5}). The statistical correlation
of this value with that given in equation~(\ref{eqn:B5}) is
\roundednum{2}{\BFsigBFeoStatCorr}. The systematic uncertainties are
fully correlated between both values. This value is consistent with
zero as is the previous measurement, by the ACCMOR collaboration, of
\DtoKppppp (excluding \PKshort, but not selecting for \Peta or
\Pomega) and has an uncertainty more than order of magnitude smaller
than that reported by ACCMOR~\cite{Barlag:1992ww}. Compared to the
hypothesis of there being no such decay, this result has a
significance of \roundednum{1}{\BFeoSignificance}. By integrating the
sampled posterior distribution, we calculate its upper limit at
\SI{95}{\!\percent} credibility is
\roundednum{1}{\BFeoUpperLimit}. Our study was not optimized for this
channel. To definitively measure it, one should conduct a dedicated
search with a \PKshort veto instead of a \PKshort selection.


\section{Conclusion}

We used \SI{988}{fb^{-1}} of data collected by the Belle experiment to
measure the branching fraction for \DtoKKppp, whose precision is
dominated by its statistical uncertainty. We do not observe
statistically significant evidence for the occurrence of the decay and
therefore report its upper limit at \SI{95}{\!\percent} credibility:
\begin{equation}
  \roundEverything{1}
  \mathcal{B}(\KKppp) < \num{\BFKKpiUpperLimit}.
\end{equation}
The Belle~II experiment aims to collect fifty times more integrated
luminosity than Belle. With such an increase of data, definitive
observation of a branching fraction as small as a few \num{e-5} will
be possible or an upper limit at order \num{e-5} can be established.

\section*{Acknowledgements}

We thank the KEKB group for the excellent operation of the
accelerator, and the KEK cryogenics group for the efficient
operation of the solenoid.

\bibliography{bibliography}

\begin{thebibliography}{30}%
\makeatletter
\providecommand \@ifxundefined [1]{%
 \@ifx{#1\undefined}
}%
\providecommand \@ifnum [1]{%
 \ifnum #1\expandafter \@firstoftwo
 \else \expandafter \@secondoftwo
 \fi
}%
\providecommand \@ifx [1]{%
 \ifx #1\expandafter \@firstoftwo
 \else \expandafter \@secondoftwo
 \fi
}%
\providecommand \natexlab [1]{#1}%
\providecommand \enquote  [1]{``#1''}%
\providecommand \bibnamefont  [1]{#1}%
\providecommand \bibfnamefont [1]{#1}%
\providecommand \citenamefont [1]{#1}%
\providecommand \href@noop [0]{\@secondoftwo}%
\providecommand \href [0]{\begingroup \@sanitize@url \@href}%
\providecommand \@href[1]{\@@startlink{#1}\@@href}%
\providecommand \@@href[1]{\endgroup#1\@@endlink}%
\providecommand \@sanitize@url [0]{\catcode `\\12\catcode `\$12\catcode
  `\&12\catcode `\#12\catcode `\^12\catcode `\_12\catcode `\%12\relax}%
\providecommand \@@startlink[1]{}%
\providecommand \@@endlink[0]{}%
\providecommand \url  [0]{\begingroup\@sanitize@url \@url }%
\providecommand \@url [1]{\endgroup\@href {#1}{\urlprefix }}%
\providecommand \urlprefix  [0]{URL }%
\providecommand \Eprint [0]{\href }%
\providecommand \doibase [0]{http://dx.doi.org/}%
\providecommand \selectlanguage [0]{\@gobble}%
\providecommand \bibinfo  [0]{\@secondoftwo}%
\providecommand \bibfield  [0]{\@secondoftwo}%
\providecommand \translation [1]{[#1]}%
\providecommand \BibitemOpen [0]{}%
\providecommand \bibitemStop [0]{}%
\providecommand \bibitemNoStop [0]{.\EOS\space}%
\providecommand \EOS [0]{\spacefactor3000\relax}%
\providecommand \BibitemShut  [1]{\csname bibitem#1\endcsname}%
\let\auto@bib@innerbib\@empty
\bibitem [{\citenamefont {Aaij}\ \emph {et~al.}(2019)\citenamefont {Aaij} \emph
  {et~al.}}]{Aaij:2019kcg}%
  \BibitemOpen
  \bibfield  {author} {\bibinfo {author} {\bibfnamefont {R.}~\bibnamefont
  {Aaij}} \emph {et~al.} (\bibinfo {collaboration} {LHCb Collaboration}),\
  }\href {\doibase 10.1103/PhysRevLett.122.211803} {\bibfield  {journal}
  {\bibinfo  {journal} {Phys. Rev. Lett.}\ }\textbf {\bibinfo {volume} {122}},\
  \bibinfo {pages} {211803} (\bibinfo {year} {2019})},\ \Eprint
  {http://arxiv.org/abs/1903.08726} {arXiv:1903.08726 [hep-ex]} \BibitemShut
  {NoStop}%
\bibitem [{\citenamefont {Grossman}\ \emph {et~al.}(2007)\citenamefont
  {Grossman}, \citenamefont {Kagan},\ and\ \citenamefont
  {Nir}}]{Grossman:2006jg}%
  \BibitemOpen
  \bibfield  {author} {\bibinfo {author} {\bibfnamefont {Y.}~\bibnamefont
  {Grossman}}, \bibinfo {author} {\bibfnamefont {A.~L.}\ \bibnamefont {Kagan}},
  \ and\ \bibinfo {author} {\bibfnamefont {Y.}~\bibnamefont {Nir}},\ }\href
  {\doibase 10.1103/PhysRevD.75.036008} {\bibfield  {journal} {\bibinfo
  {journal} {Phys. Rev. D}\ }\textbf {\bibinfo {volume} {75}},\ \bibinfo
  {pages} {036008} (\bibinfo {year} {2007})},\ \Eprint
  {http://arxiv.org/abs/hep-ph/0609178} {arXiv:hep-ph/0609178 [hep-ph]}
  \BibitemShut {NoStop}%
\bibitem [{\citenamefont {Brod}\ \emph {et~al.}(2012)\citenamefont {Brod},
  \citenamefont {Kagan},\ and\ \citenamefont {Zupan}}]{Brod:2011re}%
  \BibitemOpen
  \bibfield  {author} {\bibinfo {author} {\bibfnamefont {J.}~\bibnamefont
  {Brod}}, \bibinfo {author} {\bibfnamefont {A.~L.}\ \bibnamefont {Kagan}}, \
  and\ \bibinfo {author} {\bibfnamefont {J.}~\bibnamefont {Zupan}},\ }\href
  {\doibase 10.1103/PhysRevD.86.014023} {\bibfield  {journal} {\bibinfo
  {journal} {Phys. Rev. D}\ }\textbf {\bibinfo {volume} {86}},\ \bibinfo
  {pages} {014023} (\bibinfo {year} {2012})},\ \Eprint
  {http://arxiv.org/abs/1111.5000} {arXiv:1111.5000 [hep-ph]} \BibitemShut
  {NoStop}%
\bibitem [{\citenamefont {Inguglia}(2013)}]{lnguglia:2013mha}%
  \BibitemOpen
  \bibfield  {author} {\bibinfo {author} {\bibfnamefont {G.}~\bibnamefont
  {Inguglia}},\ }\href {\doibase 10.1142/9789814522519_0022} {\bibfield
  {journal} {\bibinfo  {journal} {Subnucl. Ser.}\ }\textbf {\bibinfo {volume}
  {49}},\ \bibinfo {pages} {407} (\bibinfo {year} {2013})},\ \Eprint
  {http://arxiv.org/abs/1109.4494} {arXiv:1109.4494 [hep-ph]} \BibitemShut
  {NoStop}%
\bibitem [{\citenamefont {Cheng}\ and\ \citenamefont
  {Chiang}(2012)}]{Cheng:2012wr}%
  \BibitemOpen
  \bibfield  {author} {\bibinfo {author} {\bibfnamefont {H.-Y.}\ \bibnamefont
  {Cheng}}\ and\ \bibinfo {author} {\bibfnamefont {C.-W.}\ \bibnamefont
  {Chiang}},\ }\href {\doibase 10.1103/PhysRevD.85.034036} {\bibfield
  {journal} {\bibinfo  {journal} {Phys. Rev. D}\ }\textbf {\bibinfo {volume}
  {85}},\ \bibinfo {pages} {034036} (\bibinfo {year} {2012})},\ \bibinfo {note}
  {[Erratum: Phys.Rev.D 85, 079903 (2012)]},\ \Eprint
  {http://arxiv.org/abs/1201.0785} {arXiv:1201.0785 [hep-ph]} \BibitemShut
  {NoStop}%
\bibitem [{\citenamefont {Bhattacharya}\ \emph {et~al.}(2012)\citenamefont
  {Bhattacharya}, \citenamefont {Gronau},\ and\ \citenamefont
  {Rosner}}]{Bhattacharya:2012ah}%
  \BibitemOpen
  \bibfield  {author} {\bibinfo {author} {\bibfnamefont {B.}~\bibnamefont
  {Bhattacharya}}, \bibinfo {author} {\bibfnamefont {M.}~\bibnamefont
  {Gronau}}, \ and\ \bibinfo {author} {\bibfnamefont {J.~L.}\ \bibnamefont
  {Rosner}},\ }\href {\doibase 10.1103/PhysRevD.85.054014} {\bibfield
  {journal} {\bibinfo  {journal} {Phys. Rev. D}\ }\textbf {\bibinfo {volume}
  {85}},\ \bibinfo {pages} {054014} (\bibinfo {year} {2012})},\ \Eprint
  {http://arxiv.org/abs/1201.2351} {arXiv:1201.2351 [hep-ph]} \BibitemShut
  {NoStop}%
\bibitem [{\citenamefont {Cheng}\ and\ \citenamefont
  {Chiang}(2019)}]{Cheng:2019ggx}%
  \BibitemOpen
  \bibfield  {author} {\bibinfo {author} {\bibfnamefont {H.-Y.}\ \bibnamefont
  {Cheng}}\ and\ \bibinfo {author} {\bibfnamefont {C.-W.}\ \bibnamefont
  {Chiang}},\ }\href {\doibase 10.1103/PhysRevD.100.093002} {\bibfield
  {journal} {\bibinfo  {journal} {Phys. Rev. D}\ }\textbf {\bibinfo {volume}
  {100}},\ \bibinfo {pages} {093002} (\bibinfo {year} {2019})},\ \Eprint
  {http://arxiv.org/abs/1909.03063} {arXiv:1909.03063 [hep-ph]} \BibitemShut
  {NoStop}%
\bibitem [{\citenamefont {Charles}\ \emph {et~al.}(2011)\citenamefont {Charles}
  \emph {et~al.}}]{Charles:2011va}%
  \BibitemOpen
  \bibfield  {author} {\bibinfo {author} {\bibfnamefont {J.}~\bibnamefont
  {Charles}} \emph {et~al.},\ }\href {\doibase 10.1103/PhysRevD.84.033005}
  {\bibfield  {journal} {\bibinfo  {journal} {Phys. Rev. D}\ }\textbf {\bibinfo
  {volume} {84}},\ \bibinfo {pages} {033005} (\bibinfo {year} {2011})},\
  \Eprint {http://arxiv.org/abs/1106.4041} {arXiv:1106.4041 [hep-ph]}
  \BibitemShut {NoStop}%
\bibitem [{\citenamefont {Altmannshofer}\ \emph {et~al.}(2012)\citenamefont
  {Altmannshofer}, \citenamefont {Primulando}, \citenamefont {Yu},\ and\
  \citenamefont {Yu}}]{Altmannshofer:2012ur}%
  \BibitemOpen
  \bibfield  {author} {\bibinfo {author} {\bibfnamefont {W.}~\bibnamefont
  {Altmannshofer}}, \bibinfo {author} {\bibfnamefont {R.}~\bibnamefont
  {Primulando}}, \bibinfo {author} {\bibfnamefont {C.-T.}\ \bibnamefont {Yu}},
  \ and\ \bibinfo {author} {\bibfnamefont {F.}~\bibnamefont {Yu}},\ }\href
  {\doibase 10.1007/JHEP04(2012)049} {\bibfield  {journal} {\bibinfo  {journal}
  {J. High Energy Phys.}\ }\textbf {\bibinfo {volume} {04}},\ \bibinfo {pages}
  {049} (\bibinfo {year} {2012})},\ \Eprint {http://arxiv.org/abs/1202.2866}
  {arXiv:1202.2866 [hep-ph]} \BibitemShut {NoStop}%
\bibitem [{\citenamefont {Giudice}\ \emph {et~al.}(2012)\citenamefont
  {Giudice}, \citenamefont {Isidori},\ and\ \citenamefont
  {Paradisi}}]{Giudice:2012qq}%
  \BibitemOpen
  \bibfield  {author} {\bibinfo {author} {\bibfnamefont {G.~F.}\ \bibnamefont
  {Giudice}}, \bibinfo {author} {\bibfnamefont {G.}~\bibnamefont {Isidori}}, \
  and\ \bibinfo {author} {\bibfnamefont {P.}~\bibnamefont {Paradisi}},\ }\href
  {\doibase 10.1007/JHEP04(2012)060} {\bibfield  {journal} {\bibinfo  {journal}
  {J. High Energy Phys.}\ }\textbf {\bibinfo {volume} {04}},\ \bibinfo {pages}
  {060} (\bibinfo {year} {2012})},\ \Eprint {http://arxiv.org/abs/1201.6204}
  {arXiv:1201.6204 [hep-ph]} \BibitemShut {NoStop}%
\bibitem [{\citenamefont {Grossman}\ \emph {et~al.}(2012)\citenamefont
  {Grossman}, \citenamefont {Kagan},\ and\ \citenamefont
  {Zupan}}]{Grossman:2012eb}%
  \BibitemOpen
  \bibfield  {author} {\bibinfo {author} {\bibfnamefont {Y.}~\bibnamefont
  {Grossman}}, \bibinfo {author} {\bibfnamefont {A.~L.}\ \bibnamefont {Kagan}},
  \ and\ \bibinfo {author} {\bibfnamefont {J.}~\bibnamefont {Zupan}},\ }\href
  {\doibase 10.1103/PhysRevD.85.114036} {\bibfield  {journal} {\bibinfo
  {journal} {Phys. Rev. D}\ }\textbf {\bibinfo {volume} {85}},\ \bibinfo
  {pages} {114036} (\bibinfo {year} {2012})},\ \Eprint
  {http://arxiv.org/abs/1204.3557} {arXiv:1204.3557 [hep-ph]} \BibitemShut
  {NoStop}%
\bibitem [{\citenamefont {Aaij}\ \emph {et~al.}(2021)\citenamefont {Aaij} \emph
  {et~al.}}]{LHCb:2021rou}%
  \BibitemOpen
  \bibfield  {author} {\bibinfo {author} {\bibfnamefont {R.}~\bibnamefont
  {Aaij}} \emph {et~al.} (\bibinfo {collaboration} {LHCb}),\ }\href {\doibase
  10.1007/JHEP06(2021)019} {\bibfield  {journal} {\bibinfo  {journal} {JHEP}\
  }\textbf {\bibinfo {volume} {06}},\ \bibinfo {pages} {019} (\bibinfo {year}
  {2021})},\ \Eprint {http://arxiv.org/abs/2103.11058} {arXiv:2103.11058
  [hep-ex]} \BibitemShut {NoStop}%
\bibitem [{\citenamefont {Abashian}\ \emph {et~al.}(2002)\citenamefont
  {Abashian} \emph {et~al.}}]{Abashian:2000cg}%
  \BibitemOpen
  \bibfield  {author} {\bibinfo {author} {\bibfnamefont {A.}~\bibnamefont
  {Abashian}} \emph {et~al.},\ }\href {\doibase 10.1016/S0168-9002(01)02013-7}
  {\bibfield  {journal} {\bibinfo  {journal} {Nucl. Instrum. Meth. A}\ }\textbf
  {\bibinfo {volume} {479}},\ \bibinfo {pages} {117} (\bibinfo {year}
  {2002})}\BibitemShut {NoStop}%
\bibitem [{\citenamefont {Brodzicka}\ \emph {et~al.}(2012)\citenamefont
  {Brodzicka} \emph {et~al.}}]{Brodzicka:2012jm}%
  \BibitemOpen
  \bibfield  {author} {\bibinfo {author} {\bibfnamefont {J.}~\bibnamefont
  {Brodzicka}} \emph {et~al.} (\bibinfo {collaboration} {Belle
  Collaboration}),\ }\href {\doibase 10.1093/ptep/pts072} {\bibfield  {journal}
  {\bibinfo  {journal} {Prog. Theor. Exp. Phys.}\ }\textbf {\bibinfo {volume}
  {2012}},\ \bibinfo {pages} {04D001} (\bibinfo {year} {2012})},\ \Eprint
  {http://arxiv.org/abs/1212.5342} {arXiv:1212.5342 [hep-ex]} \BibitemShut
  {NoStop}%
\bibitem [{\citenamefont {Kurokawa}\ and\ \citenamefont
  {Kikutani}(2003)}]{Kurokawa:2001nw}%
  \BibitemOpen
  \bibfield  {author} {\bibinfo {author} {\bibfnamefont {S.}~\bibnamefont
  {Kurokawa}}\ and\ \bibinfo {author} {\bibfnamefont {E.}~\bibnamefont
  {Kikutani}},\ }\href {\doibase 10.1016/S0168-9002(02)01771-0} {\bibfield
  {journal} {\bibinfo  {journal} {Nucl. Instrum. Meth. A}\ }\textbf {\bibinfo
  {volume} {499}},\ \bibinfo {pages} {1} (\bibinfo {year} {2003})}\BibitemShut
  {NoStop}%
\bibitem [{\citenamefont {Abe}\ \emph {et~al.}(2013)\citenamefont {Abe} \emph
  {et~al.}}]{Abe:2013kxa}%
  \BibitemOpen
  \bibfield  {author} {\bibinfo {author} {\bibfnamefont {T.}~\bibnamefont
  {Abe}} \emph {et~al.},\ }\href {\doibase 10.1093/ptep/pts102} {\bibfield
  {journal} {\bibinfo  {journal} {Prog. Theor. Exp. Phys.}\ }\textbf {\bibinfo
  {volume} {2013}},\ \bibinfo {pages} {03A001} (\bibinfo {year}
  {2013})}\BibitemShut {NoStop}%
\bibitem [{Note1()}]{Note1}%
  \BibitemOpen
  \bibinfo {note} {For brevity of notation, we denote all branching fractions
  by the final state since the initial state, \protect \PDplus , is always the
  same. All results include the charge-conjugated decays.}\BibitemShut {Stop}%
\bibitem [{\citenamefont {Link}\ \emph {et~al.}(2001)\citenamefont {Link} \emph
  {et~al.}}]{Link:2001hz}%
  \BibitemOpen
  \bibfield  {author} {\bibinfo {author} {\bibfnamefont {J.~M.}\ \bibnamefont
  {Link}} \emph {et~al.} (\bibinfo {collaboration} {FOCUS Collaboration}),\
  }\href {\doibase 10.1103/PhysRevLett.87.162001} {\bibfield  {journal}
  {\bibinfo  {journal} {Phys. Rev. Lett.}\ }\textbf {\bibinfo {volume} {87}},\
  \bibinfo {pages} {162001} (\bibinfo {year} {2001})},\ \Eprint
  {http://arxiv.org/abs/hep-ex/0105031} {arXiv:hep-ex/0105031 [hep-ex]}
  \BibitemShut {NoStop}%
\bibitem [{\citenamefont {Zyla}\ \emph {et~al.}(2020)\citenamefont {Zyla} \emph
  {et~al.}}]{PDG2020}%
  \BibitemOpen
  \bibfield  {author} {\bibinfo {author} {\bibfnamefont {P.}~\bibnamefont
  {Zyla}} \emph {et~al.} (\bibinfo {collaboration} {Particle Data Group}),\
  }\href {\doibase 10.1093/ptep/ptaa104} {\bibfield  {journal} {\bibinfo
  {journal} {Prog. Theor. Exp. Phys.}\ }\textbf {\bibinfo {volume} {2020}},\
  \bibinfo {pages} {083C01} (\bibinfo {year} {2020})}\BibitemShut {NoStop}%
\bibitem [{\citenamefont {Lange}(2001)}]{Lange:2001uf}%
  \BibitemOpen
  \bibfield  {author} {\bibinfo {author} {\bibfnamefont {D.~J.}\ \bibnamefont
  {Lange}},\ }\href {\doibase 10.1016/S0168-9002(01)00089-4} {\bibfield
  {journal} {\bibinfo  {journal} {Nucl. Instrum. Meth. A}\ }\textbf {\bibinfo
  {volume} {462}},\ \bibinfo {pages} {152} (\bibinfo {year}
  {2001})}\BibitemShut {NoStop}%
\bibitem [{\citenamefont {Brun}\ \emph {et~al.}(1987)\citenamefont {Brun},
  \citenamefont {Bruyant}, \citenamefont {Maire}, \citenamefont {McPherson},\
  and\ \citenamefont {Zanarini}}]{Brun:1987ma}%
  \BibitemOpen
  \bibfield  {author} {\bibinfo {author} {\bibfnamefont {R.}~\bibnamefont
  {Brun}}, \bibinfo {author} {\bibfnamefont {F.}~\bibnamefont {Bruyant}},
  \bibinfo {author} {\bibfnamefont {M.}~\bibnamefont {Maire}}, \bibinfo
  {author} {\bibfnamefont {A.~C.}\ \bibnamefont {McPherson}}, \ and\ \bibinfo
  {author} {\bibfnamefont {P.}~\bibnamefont {Zanarini}},\ }\href@noop {} {\
  \textbf {\bibinfo {volume} {CERN-DD-EE-84-1}} (\bibinfo {year}
  {1987})}\BibitemShut {NoStop}%
\bibitem [{\citenamefont {Fox}\ and\ \citenamefont
  {Wolfram}(1978)}]{PhysRevLett.41.1581}%
  \BibitemOpen
  \bibfield  {author} {\bibinfo {author} {\bibfnamefont {G.~C.}\ \bibnamefont
  {Fox}}\ and\ \bibinfo {author} {\bibfnamefont {S.}~\bibnamefont {Wolfram}},\
  }\href {\doibase 10.1103/PhysRevLett.41.1581} {\bibfield  {journal} {\bibinfo
   {journal} {Phys. Rev. Lett.}\ }\textbf {\bibinfo {volume} {41}},\ \bibinfo
  {pages} {1581} (\bibinfo {year} {1978})}\BibitemShut {NoStop}%
\bibitem [{\citenamefont {Feindt}\ and\ \citenamefont
  {Kerzel}(2006)}]{FEINDT2006190}%
  \BibitemOpen
  \bibfield  {author} {\bibinfo {author} {\bibfnamefont {M.}~\bibnamefont
  {Feindt}}\ and\ \bibinfo {author} {\bibfnamefont {U.}~\bibnamefont
  {Kerzel}},\ }\href {\doibase https://doi.org/10.1016/j.nima.2005.11.166}
  {\bibfield  {journal} {\bibinfo  {journal} {Nucl. Instrum. Meth. A}\ }\textbf
  {\bibinfo {volume} {559}},\ \bibinfo {pages} {190 } (\bibinfo {year}
  {2006})}\BibitemShut {NoStop}%
\bibitem [{Note2()}]{Note2}%
  \BibitemOpen
  \bibinfo {note} {We use a unit system in whcih energy, mass, and momentum all
  have units of \si {eV}.}\BibitemShut {Stop}%
\bibitem [{\citenamefont {Hoecker}\ \emph {et~al.}(2007)\citenamefont {Hoecker}
  \emph {et~al.}}]{hoecker2007tmva}%
  \BibitemOpen
  \bibfield  {author} {\bibinfo {author} {\bibfnamefont {A.}~\bibnamefont
  {Hoecker}} \emph {et~al.},\ }\href@noop {} {\bibfield  {journal} {\bibinfo
  {journal} {arXiv:physics/0703039}\ } (\bibinfo {year} {2007})}\BibitemShut
  {NoStop}%
\bibitem [{\citenamefont {Levit}(2019)}]{Levit:2019jqe}%
  \BibitemOpen
  \bibfield  {author} {\bibinfo {author} {\bibfnamefont {D.}~\bibnamefont
  {Levit}},\ }\href {https://mediatum.ub.tum.de/?id=1468810} {\ \textbf
  {\bibinfo {volume} {BELLE2-PTHESIS-2021-017}} (\bibinfo {year}
  {2019})}\BibitemShut {NoStop}%
\bibitem [{\citenamefont {Barlag}\ \emph {et~al.}(1992)\citenamefont {Barlag}
  \emph {et~al.}}]{Barlag:1992ww}%
  \BibitemOpen
  \bibfield  {author} {\bibinfo {author} {\bibfnamefont {S.}~\bibnamefont
  {Barlag}} \emph {et~al.} (\bibinfo {collaboration} {ACCMOR Collaboration}),\
  }\href {\doibase 10.1007/BF01565095} {\bibfield  {journal} {\bibinfo
  {journal} {Z. Phys. C}\ }\textbf {\bibinfo {volume} {55}},\ \bibinfo {pages}
  {383} (\bibinfo {year} {1992})}\BibitemShut {NoStop}%
\bibitem [{\citenamefont {Caldwell}\ \emph {et~al.}(2009)\citenamefont
  {Caldwell} \emph {et~al.}}]{caldwell2009bat}%
  \BibitemOpen
  \bibfield  {author} {\bibinfo {author} {\bibfnamefont {A.}~\bibnamefont
  {Caldwell}} \emph {et~al.},\ }\href@noop {} {\bibfield  {journal} {\bibinfo
  {journal} {Comp. Phys. Commun.}\ }\textbf {\bibinfo {volume} {180}},\
  \bibinfo {pages} {2197} (\bibinfo {year} {2009})}\BibitemShut {NoStop}%
\bibitem [{\citenamefont {Beaujean}\ \emph {et~al.}(2018)\citenamefont
  {Beaujean}, \citenamefont {Caldwell}, \citenamefont {Greenwald},
  \citenamefont {Kröninger},\ and\ \citenamefont
  {Schulz}}]{beaujean_frederik_2018_1322675}%
  \BibitemOpen
  \bibfield  {author} {\bibinfo {author} {\bibfnamefont {F.}~\bibnamefont
  {Beaujean}}, \bibinfo {author} {\bibfnamefont {A.}~\bibnamefont {Caldwell}},
  \bibinfo {author} {\bibfnamefont {D.}~\bibnamefont {Greenwald}}, \bibinfo
  {author} {\bibfnamefont {K.}~\bibnamefont {Kröninger}}, \ and\ \bibinfo
  {author} {\bibfnamefont {O.}~\bibnamefont {Schulz}},\ }\href {\doibase
  10.5281/zenodo.1322675} {\enquote {\bibinfo {title} {{BAT release, version
  1.0.0}},}\ } (\bibinfo {year} {2018})\BibitemShut {NoStop}%
\bibitem [{\citenamefont {Skwarnicki}(1986)}]{Skwarnicki:1986xj}%
  \BibitemOpen
  \bibfield  {author} {\bibinfo {author} {\bibfnamefont {T.}~\bibnamefont
  {Skwarnicki}},\ }\href@noop {} {\ \textbf {\bibinfo {volume}
  {DESY-F31-86-02}} (\bibinfo {year} {1986})}\BibitemShut {NoStop}%
\end{thebibliography}%

\end{document}